\documentclass[reqno]{amsart}

\usepackage{amsfonts}

\usepackage{amssymb}

\usepackage{amsmath}

\usepackage{amsthm}

\numberwithin{equation}{section}

\newcommand{\Space}[2]{ \mathbb{#1}^{#2} }

\newcommand{\ub}[3][]{\left\{\!#1\left[#2,#3\right]\!#1\right\}}

\newcommand{\horb}{\mathcal{O}_{\hbar}}

\newtheorem{Corollary}{Corollary}[section]

\newtheorem{Thm}{Theorem}[section]

\newtheorem{Defn}{Definition}[section]

\newtheorem{Lemma}{Lemma}[section]

\newcommand{\Partial}[1]{ \frac{\partial}{\partial #1} }

\newcommand{\Fracpartial}[2]{\frac{\partial #1}{\partial #2} }

\newcommand{\Diffl}[1]{\frac{d}{d #1}}

\newcommand{\Fracdiffl}[2]{\frac{d #1}{d #2}}

\newcommand{\Lalg}[2]{ \mathfrak{#1}^{#2} }

\newcommand{\Heisn}{\Space{H}{n}}

\newcommand{\heisn}{\mathfrak{h}_{n}}

\newcommand{\heisnstar}{\Lalg{h}{*}_{n}}

\newcommand{\uorb}{\mathcal{O}_{h}}

\newcommand{\hilbh}{\mathcal{H}_{h}}

\newcommand{\iilbh}{\mathcal{I}_{h}}

\newcommand{\lkerh}{\mathcal{L}_{h}}

\newcommand{\lkero}{\mathcal{L}_{0}}

\newcommand{\loneh}{L^1(\Heisn)}

\newcommand{\ltwoh}{L^2(\Heisn)}

\newcommand{\ltwohnstar}{L^2 (\heisnstar) }

\newcommand{\fock}{F^2(\uorb)}

\newcommand{\antid}{\mathcal{A}}

\newcommand{\zerodel}{\delta(s) \delta(x) \delta(y) }

\newcommand{\zerodelsone}{\delta^{(1)}(s) \delta(x) \delta(y) }

\newcommand{\zerodelxone}{\delta(s) \delta^{(1)}(x) \delta(y) }

\newcommand{\zerodelyone}{\delta(s) \delta(x) \delta^{(1)}(y) }

\newcommand{\zerodelxtwo}{\delta(s) \delta^{(2)}(x) \delta(y) }

\newcommand{\zerodelytwo}{\delta(s) \delta(x) \delta^{(2)}(y) }

\newcommand{\fort}{\mathcal{F}}

\newcommand{\vab}{v_{(h,a,b)}}

\newcommand{\voo}{v_{(h,0,0)}}

\newcommand{\lab}{l_{(h,a,b)}}

\newcommand{\loo}{l_{(h,0,0)}}

\newcommand{\labo}{l_{(0,a,b)}}

\newcommand{\qporb}{\mathcal{O}_{(q,p)}}

\newcommand{\statem}{\mathcal{S}_h}

\newcommand{\statemo}{\mathcal{S}_0}

\begin{document}

\title{Classical and Quantum Coherent States}

\author{Alastair Brodlie}

\date{12/03/2003}

\address{%
School of Mathematics\\
University of Leeds\\
Leeds LS2\,9JT\\
UK}

\email{abrodlie@amsta.leeds.ac.uk}

\maketitle

\begin{abstract}
p-Mechanics is a consistent physical theory which describes both quantum and classical mechanics simultaneously \cite{Kisil02.1,Kisil02}. We continue the development of p-mechanics by introducing the concept of states. The set of coherent states we introduce allow us to evaluate classical observables at any point of phase space simultaneously to evaluating quantum probability amplitudes. The example of the forced harmonic oscilator is used to demonstrate these concepts.
\end{abstract}

{\bf Keywords:} Heisenberg group, quantum mechanics, classical mechanics, coherent states, pure states, forced harmonic oscillator, $C^*$-algebras, kernels, scattering matrix, statistical mechanics.

{\bf 2000 Mathematics Subject Classification:} Primary 81R05; Secondary 81R30, 22E27, 22E70, 43A65.

\tableofcontents

\section{Introduction}

In this paper we continue the development of p-mechanics \cite{Kisil02} ,\cite{Kisil02.1}. p-mechanics is  a consistent physical theory which simultaneously describes both quantum and classical mechanics. It uses the representation theory of the Heisenberg group to show that both quantum and classical mechanics are derived from the same source.

In this paper we introduce the concept of states to p-mechanics. These
are defined in subsection \ref{subsect:whatarestates} as functionals on the $C^*$-algebra of observables which come in two equivalent forms: as elements of a Hilbert Space and as kernels on the Heisenberg group. Their time evolution is defined in subsection \ref{subsect:timeevolofstates} and it is shown that the Schr\"odinger and Heisenberg pictures are equivalent in p-mechanics.
In subsection \ref{sect:cstates} we introduce an overcomplete system of  coherent states for p-mechanics whose classical limit corresponds to the classical pure states. We introduce two forms of functional since both have their own advantages. The Hilbert space functionals are useful for deriving quantum properties of a system, while the kernels have a clearer time evolution and classical limit.

Finally in section \ref{sect:forcedosc} we apply this theory to the example of the forced oscillator. It is shown that both the quantum and classical pictures are derived from the same source. The important features of the classical case are proved using p-mechanics in subsections \ref{sect:forcedoscclass} and \ref{sect:periodandres}. Some of the features of the quantum case are proved in subsection \ref{sect:forcedoscquant}.

\section{Notation and Preliminaries}

\subsection{Notation}

$I$ is the identity $n$ by $n$ matrix. \newline
$\hat{f}$ is the Fourier Transform of a function $f$ (e.g. $\hat{f}(x)=
\int_{\Space{R}{n}}f(q) e^{-2 \pi i q.x}dq$ ). We also use $\fort$ to denote the Fourier Transform when this form is more convenient (e.g. $\fort_{(x,y)} (f(q,p)) = \int_{\Space{R}{n}}f(q,p) e^{-2 \pi i (q.x+p.y)} dq dp)$ .

\subsection{Preliminaries}
In this section we give a very brief overview of \cite{Kisil02.1}, which provides an introduction to p-mechanics.
At the heart of this paper is the Heisenberg group (\cite{Folland89}, \cite{Taylor86}).
\begin{Defn}
The Heisenberg Group (denoted $\Space{H}{n}$) is the set of all triples in $\Space{R}{} \times \Space{R}{n} \times \Space{R}{n}$ under the law of multiplication
\begin{equation} \label{heismult}
(s,x,y) * (s',x',y') = (s+s' + \frac{1}{2} (x \cdot y' -x' \cdot y),x+x',y+y').
\end{equation}
\end{Defn}
The non-commutative convolution of two functions $B_1,B_2$ defined on $\Heisn$ is
\begin{equation} \nonumber
(B_1 * B_2)(g) = \int_{\Heisn} B_1 (h) B_2 (h^{-1} g) dh = \int_{\Heisn} B_1 (g h^{-1}) B_2 (h) dh,
\end{equation}
where $dh$ is Harr Measure on $\Heisn$ which is just the Lebesgue measure $ds \, dx \, dy$. This can be extended to distributions in a natural way \cite{KirillovGvishiani82}. The Lie Algebra $\heisn$ can be realised by the left invariant vector fields
\begin{displaymath}
\begin{array}{ccc}
S=\Partial{s}, & X_j = \Partial{x_j} - \frac{y_j}{2} \Partial{s}, & Y_j = \Partial{y_j} + \frac{x_j}{2} \Partial{s},
\end{array}
\end{displaymath}
with the Heisenberg commutator relations
\begin{equation} \nonumber
[ X_i , Y_j ] = \delta_{ij} S.
\end{equation}
The dual space to the Lie Algebra $\heisn^*$ is spanned by the left invariant first order differential forms.
One of the principal ways of transfering between $\ltwoh$ and $\ltwohnstar$ is by the Fourier transform on $\Heisn$ \cite{Kirillov99}
\begin{equation} \label{eq:liegpfourier}
\hat{\phi}(F) = \int_{\heisn} \phi (\exp X) e^{-2 \pi i \langle X,F \rangle } \, dX.
\end{equation}
For the Heisenberg group this has the simple form
\begin{equation} \nonumber
\hat{\phi}(h,q,p) = \int_{\Space{R}{2n+1}} \phi (s,x,y) e^{-2\pi i (hs+q.x+p.y)} ds dx dy
\end{equation}
which is just the usual Fourier transform on $\Space{R}{2n+1}$.
Kirillov's Method of Orbits is of great importance in the p-mechanical construction. For a discussion of the Method of Orbits, see \cite{Kirillov99} or \cite[Chap 15]{Kirillov76}; its relation to p-mechanics is described in \cite{Kisil02.1}. The Method of Orbits  uses the representaion $Ad^*$ of $\Heisn$ on $\heisn^*$, known as the coadjoint representation. The exact form of this representation for $\Heisn$ is
\begin{equation} \nonumber
Ad^* (s,x,y): (h,q,p) \mapsto (h,q+hy,p-hx).
\end{equation}
Note here that $(s,x,y) \in \Heisn$ and $(h,q,p) \in \heisn^* $ --- this choice of letters will be used throughout this paper. The orbits of $Ad^*$ come in two forms, Euclidean spaces $\Space{R}{2n}$ and singleton points
\begin{eqnarray} \nonumber
\uorb &=& \{ (h,q,p) : \textrm{ fixed } h \neq 0  , \textrm{ and  any } q,p \in \Space{R}{n} \} ; \\ \nonumber
\qporb &=& \{ (0,q,p) : \textrm{  any   } q,p \in \Space{R}{n} \} .
\end{eqnarray}
We now introduce $\fock$ which is a subspace of $L^2 (\uorb )$, we use this space because it is irreducible under the representation $Ad^*$
\begin{equation} \nonumber
\fock = \{ f_h (q,p) \in L^2 (\uorb)  : D^j_h f_h =0 ,\textrm{ for } 1 \leq j \leq n \} ,
\end{equation}
where the operator $D_h^j$ on $L^2 (\uorb )$ is defined as
$\frac{h}{2} \left( \Partial{p_j} + c_i i \Partial{q_j} \right) + 2 \pi (c_i p_j +i q_j)$.
The inner product on $\fock$ is given by
\begin{equation} \label{eq:fockip}
\langle v_1, v_2 \rangle_{\fock} = \left( \frac{4}{ h } \right)^{n} \int_{\Space{R}{2n}} v_1 (q,p) \overline{v_2 (q,p)} \, dq \, dp
\end{equation}
The representation $\rho_h$ of $\Heisn$ on $F^2 (\uorb)$ is provided by
\begin{equation} \label{eq:infdimrep}
\rho_h (s,x,y): f_h (q,p) \mapsto e^{-2 \pi i (hs + qx +py)} f_h(q-\frac{h}{2} y, p+\frac{h}{2} x),
\end{equation}
which is unitary with respect to the inner product defined in (\ref{eq:fockip})
The crucial theorem which motivates the whole of p-mechanics is
\begin{Thm} \nonumber
(The Stone-von Neumann Theorem) All unitary irreducible representations of the Heisenberg group, $\Space{H}{n}$, up to unitary equivalence, are either:

(i) of the form $\rho_h$ on $\fock$ from equation (\ref{eq:infdimrep}), or

(ii) for $(q,p) \in \Space{R}{2n}$ the commutative  one-dimensional representations on $\Space{C}{} = L^2 (\qporb )$
\begin{equation} \label{eq:1drep}
\rho_{(q,p)} (s,x,y)u = e^{- 2 \pi i(q.x + p.y)} u.
\end{equation}
\end{Thm}
\begin{proof}
For a proof see either \cite{Folland89} or \cite{Taylor86}.
\end{proof}
We can extend this to the representation of a function, $B$, on $\Heisn$ by
\begin{equation} \nonumber
\rho (B) = \int_{\Heisn} B(g) \rho(g) dg.
\end{equation}
The representation of distributions is done in the natural way \cite[Chap 0, Eq 3.4]{Taylor86}. The basic idea of p-mechanics is to choose particular functions or distributions on $\Heisn$ which under the infinite dimensional representation will give quantum mechanical observables while under the one dimensional representation will give classical mechanical observables. The observables are in fact operators on $L^2 (\Heisn)$ generated by convolutions of the chosen functions or distributions (more general operators on $\ltwoh$ are in use for string-like versions of p-mechanics but nothing has been published on this yet). Since $L^2(\Heisn)$ is a Hilbert space we have that the set of observables is a C*-algebra \cite{Arveson76,Dixmier77}. In doing this it is shown that both mechanics are derived from the same source. The dynamics of a physical system evolves in the p-mechanical picture using the universal brackets \cite{Kisil02}; these are defined on two observables $B_1 , B_2$ by
\begin{equation} \nonumber
\ub{B_1}{B_2}{} = \antid ( B_1 * B_2 - B_2 * B_1 ),
\end{equation}
where the operator $\antid$ is defined on exponents as (recall that $S=\Partial{s}$)
\begin{equation}
  \label{eq:defnofantid}
 S\antid=4\pi^2 I, \qquad \textrm{ where }\quad
  \antid e^{ 2\pi i h s}=\left\{
    \begin{array}{ll}
      \displaystyle
      \frac{2\pi}{i h \strut} e^{2\pi i h s}, & \textrm{if }
h \neq 0,\\
      4\pi^2 s\strut, & \textrm{if } h=0.
    \end{array}
    \right.
\end{equation}
and can be extended by linearity to the whole of $\loneh$. $\antid$ is the antiderivative operator since it is the right inverse to $\Partial{s}$. It is proved in \cite{Kisil02} that the universal brackets satisfy both the Liebniz and Jacobi identities along with being anticommutative.
For a p-mechanical system with energy $B_H$, these brackets give us a p-dynamic equation for an observable $B$:
\begin{equation} \label{eq:pdyneqn}
\frac{d B}{dt} = \ub{B}{B_H}{}.
\end{equation}

Finally we state an equation from \cite{Kisil02.1} which will be of use throughout the paper. If we define the operator $\lambda_l (g)$ for each $g \in \Heisn$ on $L^2 (\Heisn)$ as
\begin{equation} \label{eq:leftshift}
\lambda_l (g):f(h) \mapsto f(g^{-1} h)
\end{equation}
(i.e. the left regular representation) then we have the following relation
\begin{equation} \label{eq:ftandshift}
\lambda_l (g) \fort = \fort \rho_h (g).
 \end{equation}
We can alternatively write the convolution of two functions on the Heisenberg group as
\begin{equation} \label{eq:altconvol}
B_1*B_2 (g) = \int_{\Heisn} B_1 (h) \lambda_l (h) dh B_2 (g).
\end{equation}

\section{States and the Pictures of p-Mechanics}
\subsection{States} \label{subsect:whatarestates}

In this section we introduce states to p-mechanics --- these are positive linear functionals on the C*-algebra \cite{Arveson76,Dixmier77} of p-mechanical observables. For each $h \neq 0$ (the quantum case) we give two equivalent forms of states: the first form we give is as elements of a Hilbert space, the second is as integration with an apropriate kernel. For $h=0$ (the classical case) we have only one form of states, that is as integration with an apropriate kernel.

\begin{Defn}

The Hilbert space $\hilbh$, $h \in \Space{R}{} \setminus \{ 0 \}$, is the subset of functions on $\Heisn$  defined by
\begin{equation} \label{hh}
\hilbh = \left\{ e^{2 \pi ihs} f (x,y) : E^j_h f = 0 \hspace{1cm} 1 \leq j \leq n \right\}
\end{equation}
where the operator $E^j_h = \frac{h}{2} (y_j + i c_i x_j)I + 2 \pi (c_i \Partial{y_j}+ i \Partial{x_j})$ (this is the Fourier transform of $D_h^j$). The inner product on $\hilbh$ is defined as

\begin{equation}  \label{hhip}
\langle v_1 , v_2 \rangle_{\hilbh} = \left( \frac{4}{h} \right)^{n} \int_{\Space{R}{2n}} v_1 (s,x,y) \overline{v}_2 (s,x,y) \, dx \, dy.
\end{equation}
\end{Defn}
Note in equation ($\ref{hhip}$) there is no integration over the $s$ variable since for any two functions $v_1 = e^{2 \pi ihs} f_1 (x,y)$ and $v_2 = e^{2 \pi ihs} f_2 (x,y)$ in $\hilbh$
\begin{equation} \nonumber
\langle v_1 , v_2 \rangle = \int_{\Space{R}{2n}} e^{2 \pi ihs} e^{- 2 \pi ihs} f_1 (x,y) \bar{f}_2 (x,y) \, dx \, dy = \int_{\Space{R}{2n}} f_1 (x,y) \bar{f}_2 (x,y) \, dx \, dy
\end{equation}
and hence there is no $s$-dependence. It is important to note that each $\hilbh$ is invariant under convolutions.
Since the Fourier transform intertwines multiplication and differentiation we have
\begin{equation} \label{hh}
\hilbh = \left\{ e^{2 \pi ihs} \fort_{(x,y)} (f(q,p)) : f \in F^2(\horb) \right\}.
\end{equation}
$\hilbh$ is mapped into another Hilbert Space $\iilbh$ by the Fourier transform. This Hilbert Space $\iilbh$ is
\begin{equation} \nonumber
\iilbh = \left\{ j(h',q,p) = \delta(h'-h) f(q,p) : f \in \fock \right\},
\end{equation}
where $\delta$ is the Dirac delta distribution. The inner product for $j_1(h',q,p) = \delta(h'-h) f_1(q,p)$ and  $j_2(h',q,p) = \delta(h'-h) f_2(q,p)$ in $\iilbh$ is
\begin{equation} \nonumber
\langle j_1 , j_2 \rangle_{\iilbh} = \left( \frac{4}{h} \right)^n \int_{\Space{R}{2n+1}} j_1 (h',q,p) \overline{ j_2 (h',q,p) } \, dh' \, dq  \, dp = \langle f_1 , f_2 \rangle_{\fock}.
\end{equation}
We define a set of states for each $h \neq 0$ using $\hilbh$ (later in this section we will define a set of states for $h\neq0$ which are defined using a kernel and a set of states for $h=0$ by a kernel).
\begin{Defn}
If $B$ is a p-mechanical observable and $v\in \hilbh$ the p-mechanical state corresponding to $v$ is
\begin{equation} \nonumber
\langle B * v , v \rangle_{\hilbh}.
\end{equation}
\end{Defn}
In \cite{Kisil02.1} it is stated that if $A$ is a quantum mechanical observable (that is an operator on $\fock$) the state corresponding to $f \in \fock$ is
\begin{equation} \nonumber
\langle A f,f \rangle_{\fock} .
\end{equation}
We now introduce a map $\statem$ which maps vectors in $\fock$ to vectors in $\hilbh$
\begin{equation} \label{eq:statemapforhnonzero}
\statem (f(q,p)) = e^{2 \pi ihs} \hat{f}(x,y).
\end{equation}
The following Theorem proves that the states corresponding to vectors $f$ and $\statem f$ give the same expectation values for observables $B$ and $\rho_h (B)$ respectively. Before proving this we state a short Lemma which is needed to prove this Theorem.
\begin{Lemma}
The map $\statem:\hilbh \rightarrow \iilbh$ is an isometry.
\end{Lemma}

\begin{proof}
Let $v_1,v_2 \in \hilbh$ be of the form
\begin{eqnarray} \label{eq:whatarethestates}
v_1 (s,x,y) &=& \statem f_1 = e^{2 \pi ihs} \hat{f}_1 (x,y) \\ \nonumber
v_2 (s,x,y) &=& \statem f_2 = e^{2 \pi ihs} \hat{f}_2 (x,y)
\end{eqnarray}
where $f_1$ and $f_2$ are in $\fock$. Showing that $\statem$ is an isometry is equivalent to proving the relation
\begin{equation} \label{eq:plancheralpolarz}
\langle v_1,v_2 \rangle_{\hilbh} = \langle \hat{v}_1 , \hat{v}_2 \rangle_{\iilbh}
\end{equation}
where  $\hat{}$  signifies the Fourier transform on the Heisenberg group defined in equation (\ref{eq:liegpfourier}). By an elementary calculation
\begin{eqnarray} \label{eq:fourierofthetwostates}
\hat{v}_1(h',q,p) &=& \delta (h'-h) f_1(q,p) \\ \nonumber
\hat{v}_2(h',q,p) &=& \delta (h'-h) f_2(q,p).
\end{eqnarray}
From the Plancheral formula on $\Space{R}{2n}$ and the polarization identity \cite{Kirillov76} we have
\begin{equation} \nonumber
\langle f_1 , f_2 \rangle_{L^2(\Space{R}{2n})} = \langle \hat{f}_1,\hat{f}_2 \rangle_{L^2(\Space{R}{2n})} .
\end{equation}
Using this result we get
\begin{eqnarray} \nonumber
\langle v_1 , v_2 \rangle_{\hilbh} &=& \int_{\Space{R}{2n}} f_1 (x,y) \overline{f_2 (x,y)} \, dx \, dy \\ \nonumber
&=& \int_{\Space{R}{2n}} \hat{f}_1 (q,p) \overline{\hat{f}_2 (q,p)} \, dq \, dp \\ \nonumber
&=& \int_{\Space{R}{2n+1}} \delta (h'-h) \hat{f}_1 (q,p) \overline{\delta (h'-h) \hat{f}_2 (q,p)} \, dh' \, dq \, dp.
\end{eqnarray}
Then by (\ref{eq:fourierofthetwostates}) this gives us
\begin{equation} \nonumber
\langle v_1 , v_2 \rangle_{\hilbh} = \langle \hat{v}_1,\hat{v}_2 \rangle_{\iilbh}.
\end{equation}
\end{proof}

\begin{Thm} \label{hilbh-f2h}

For any  observable $B$ and any $v_1, v_2  \in \hilbh$, $h \in \Space{R}{} \setminus \{ 0 \}$, of the form given in (\ref{eq:whatarethestates}) we have the relationship
\begin{equation} \label{eq:statesrelation}
\langle B * v_1 , v_2 \rangle_{\hilbh} = \langle \rho_{h} (B) f_1, f_2 \rangle_{\fock}.
\end{equation}
\end{Thm}

\begin{proof}

From equation (\ref{eq:plancheralpolarz}) we have
\begin{equation} \label{eq:relhhih}
\langle B * v_1,v_2 \rangle_{\hilbh} = \langle \widehat{B * v_1}, \hat{v}_2 \rangle_{\iilbh}
\end{equation}
where again $\hspace{0.3cm}  \hat{} \hspace{0.3cm}$  is the Fourier transform on the Heisenberg group as described in equation (\ref{eq:liegpfourier}). Using (\ref{eq:altconvol}) equation (\ref{eq:relhhih}) can be written as
\begin{equation} \label{eq:secondrelhhih}
\langle B * v_1,v_2 \rangle_{\hilbh} = \langle \int \widehat{B (g) \lambda_l (g) dg v_1} , \hat{v}_2 \rangle_{\iilbh}.
\end{equation}
Using (\ref{eq:ftandshift}) equation (\ref{eq:secondrelhhih}) becomes
\begin{eqnarray} \nonumber
\langle B * v_1 , v_2 \rangle &=& \langle \int B (g) \rho_h (g) dg \hat{v}_1 , \hat{v}_2 \rangle \\ \nonumber
&=& \int \rho_h (B) \delta (h'-h) f_1(q,p) \overline{\delta (h'-h) f_2(q,p)} \, dq \, dp \, dh' \\ \nonumber
&=& \int \rho_h (B) f_1(q,p) \overline{f_2} (q,p) \, dq \, dp.
\end{eqnarray}
Hence the result has been proved.
\end{proof}
Taking $v_1=v_2$ in (\ref{eq:statesrelation}) shows that the states corresponding to $f$ and $\statem f$ will give the same expectation values for $\rho_h (B)$ and $B$ respectively. If we take $B$ to be a time development operator we can get probability amplitudes between states $v_1 \neq v_2$.

We now show that each of these states can also be realised by an appropriate kernel.
\begin{Thm}
If $l(s,x,y)$ is defined to be the kernel
\begin{equation} \label{eq:relationbetweenkernelandvector}
l(s,x,y) = \left(\frac{4}{h}\right)^n \int_{\Space{R}{2n}}  v((s,x,y)^{-1} (s',x',y')) \overline{v((s',x',y'))} \, dx' \, dy'.
\end{equation}
then
\begin{equation} \nonumber
\langle B*v,v \rangle_{\hilbh} = \int_{\Heisn} B(s,x,y) l(s,x,y) \, ds \, dx \, dy
\end{equation}
\end{Thm}
\begin{proof}
It is easily seen that
\begin{eqnarray} \nonumber
\lefteqn{\langle B*v,v \rangle} \\ \nonumber
&=& \left(\frac{4}{h}\right)^n \int_{\Space{R}{2n}} \int_{\Heisn} B((s,x,y)) v((s,x,y)^{-1} (s',x',y')) \\ \nonumber
&& \qquad \qquad \qquad \qquad \times \overline{v((s',x',y'))} \, ds \, dx \, dy \, dx' \, dy' \\ \label{eq:vandlsxy}
&=& \left(\frac{4}{h}\right)^n \int_{\Heisn} B((s,x,y)) \\ \nonumber
&& \qquad \qquad \times \left( \int_{\Space{R}{2n}} v((s,x,y)^{-1} (s',x',y')  )\overline{v((s',x',y'))}  \,dx' \, dy' \right) \,ds \, dx \, dy
\end{eqnarray}
Note that there is no integration over $s'$ by the definition of the $\hilbh$ inner product.
\end{proof}
\begin{Defn}
We denote the set of kernels corresponding to the elements in $\hilbh$ as $\lkerh$.
\end{Defn}

Now we introduce p-mechanical $(q,p)$ states which correspond to classical states, they are again functionals on the C*-algebra of p-mechanical observables. Pure states in classical mechanics evaluate observables at particular points of phase space, they can be realised as kernels
$\delta(q-a , p-b)$ for fixed $a,b$ in phase space, that is
\begin{equation} \label{eq:classicalpurestateeval}
\int_{\Space{R}{2n}} F(q,p) \delta(q-a , p-b) dq dp = F(a,b).
\end{equation}
We now give the p-mechanical equivalent of pure classical states.
\begin{Defn}
p-Mechanical $(q,p)$ pure states are defined to be the set of functionals, $k^0_{(a,b)}$, for fixed $a,b \in \Space{R}{2n}$ which act on observables by
\begin{equation} \label{eq:pclasspurestates}
k_{(0,a,b)}(B(s,x,y)) = \int_{\Heisn} B(s,x,y) e^{-2\pi i (a.x+b.y)}\, dx \, dy.
\end{equation}
Each $(q,p)$ pure state $k_{(0,a,b)}$ is defined entirely by its kernel $l_{(0,a,b)}$
\begin{equation}
l_{(0,a,b)} = e^{-2\pi i (a.x+b.y)}.
\end{equation}
\end{Defn}
By the definition of p-mechanisation (this is the map from classical observables to p-mechanical observables which is the inverse Fourier transform followed by the tensoring with a dirac delta function in the $s$ variable \cite[Sect. 3.3]{Kisil02.1})
\begin{equation} \label{eq:pmechanicalpurestateeval}
\int_{\Heisn} B(s,x,y) e^{-2\pi i (a.x+b.y)}\, ds \, dx \, dy = F(a,b)
\end{equation}
where $F$ is the classical observable corresponding to $B$, hence when we apply state $k_{(0,a,b)}$ to a p-mechanical observable we get the value of its classical counterpart at the point $(a,b)$ of phase space.
We introduce the map $\statemo$ which maps classical pure state kernels to p-mechanical $(q,p)$ pure state kernels
\begin{equation} \nonumber
\statemo (\xi(q,p)) = \hat{\xi} (x,y).
\end{equation}
This equation is almost identical to the relation in equation (\ref{eq:statemapforhnonzero}). The kernels $\labo$, are the Fourier transforms of the delta functions
\newline
$\delta(q-a,p-b)$, hence pure $(q,p)$ states are just the image of pure classical states.

Mixed states, as used in statistical mechanics \cite{Honerkamp98}, are linear combinations of pure states. In p-mechanics $(q,p)$ mixed states are defined in the same way.
\begin{Defn}
Define $\lkero$, to be the space of all linear combinations of $(q,p)$ pure state kernels $l_{(0,a,b)}$, that is the set of all kernels corresponding to $(q,p)$ mixed states.
\end{Defn}
The map $\statemo$ exhibits the same relations on mixed states as pure states due to the linearity of the Fourier transform.

\subsection{Time Evolution of States} \label{subsect:timeevolofstates}
We now go on to show how p-mechanical states evolve with time.  We first show how the elements of $\lkerh$, for all $h\in\Space{R}{}$ evolve with time and that this time evolution agrees with the time evolution of p-observables. In doing this we show that for the particular case of $\lkero$ the time evolution is the same as classical states under the Liouville equation. Then we show how the elements of $\hilbh$ evolve with time and prove that they agree with the Schr\"odinger picture of motion in quantum  mechanics. Before we can do any of this we need to give the definition of a Hermitian convolution.

\begin{Defn}
We call a p-mechanical observable B Hermitian if it corresponds to a Hermitian convolution, that is for any functions $F_1,F_2$ on the Heisenberg group
\begin{equation} \nonumber
\int_{\Heisn} (B*F_1)(g) \overline{F_2(g)} dg = \int_{\Heisn} F_1(g)\overline{(B*F_2)(g)} dg.
\end{equation}
\end{Defn}
If a p-observable $B$ is Hermitian then  $ B(g) = \overline{B(g^{-1})}$, this is the result of a trivial calculation. From now on we denote $\overline {B (g^{-1})}$ as $B^*$. For our purposes we just need to assume that the distribution or function, $B$, corresponding to the observable is real and  $B(s,x,y)=B(-s,-x,-y)$.

\begin{Defn}
If we have a system with energy $B_H$ then an arbitary kernel $l \in \lkerh$, $h\in \Space{R}{}$, evolves under the equation
\begin{equation} \label{eq:zerostatestimeeq}
\Fracdiffl{l}{t} =  \ub{B_H}{l}{}.
\end{equation}
\end{Defn}
We now show the time evolution of these kernels coincides with the time evolution of p-mechanical observables.
\begin{Thm} \label{Thm:timeevolofkernelisok}
If $l$ is a kernel evolving under equation (\ref{eq:zerostatestimeeq}) then for any observable $B$
\begin{equation} \nonumber
\Diffl{t} \int_{\Heisn} B \, l \, dg = \int_{\Heisn} \ub{B}{B_H}{} \, l \, dg
\end{equation}
\end{Thm}
\begin{proof}
This result can be verified by the direct calculation,
\begin{eqnarray} \nonumber
\lefteqn{\Diffl{t} \int_{\Heisn} B(s,x,y) l(s,x,y) \, ds \, dx \, dy } \\ \nonumber
&=& \int_{\Heisn} B(s,x,y) \antid (B_H *l - l *B_H ) (s,x,y) \, ds \, dx \, dy \\ \label{eq:usedintbypartsintime}
&=& - \int_{\Heisn} \antid B(s,x,y) (B_H * l - l *B_H ) (s,x,y) \, ds \, dx \, dy \\ \nonumber
&=& \int_{\Heisn} \antid ((B * B_H)(s,x,y) l (s,x,y) \\  \label{eq:usedhermitianinzero}
&& \qquad -(B_H * B)(s,x,y) l (s,x,y) ) \, ds \, dx \, dy \\ \nonumber
&=& \int_{\Heisn} \ub{B}{B_H}{}(s,x,y) l(s,x,y) \, ds \, dx \, dy
\end{eqnarray}
At (\ref{eq:usedintbypartsintime}) we have used integration by parts while (\ref{eq:usedhermitianinzero}) follows since $B_H$ is Hermitian.
\end{proof}
If we take the representation $\rho_{(q,p)}$ of equation  (\ref{eq:zerostatestimeeq}) we get the Liouville equation \cite[Eq. 5.42]{Honerkamp98} for a kernel $\statemo^{-1}(l)$ moving in a system with energy $\rho_{(q,p)} (B_H)$. This only holds for elements in $\lkero$ and can be verified by a similar calculation to \cite[Propn. 3.5]{Kisil02}.

Now we show how the vectors in $\hilbh$ evolve with time. Initially we extend our definition of $\antid$ which was initially introduced in equation (\ref{eq:defnofantid}). $\antid$ can also be defined as an operator on each $\hilbh$, $h \in \Space{R}{} \setminus \{ 0 \}$, $\antid:\hilbh \mapsto \hilbh$ by
\begin{equation} \nonumber
\antid v = \frac{2\pi}{ih} v.
\end{equation}
The adjoint of $\antid$ is $-\antid$ on each $\hilbh$, $h \in \Space{R}{} \setminus \{ 0 \}$.
\begin{Defn}
If we have a system with energy $B_H$ then an arbitrary vector $v \in \hilbh$ evolves under the equation
\begin{equation} \label{eq:timevolinhh}
\Fracdiffl{v}{t} = \antid B_H *v = B_H * \antid v
\end{equation}
\end{Defn}
The operation of left convolution preserves each $\hilbh$ so this time evolution is well defined. Equation (\ref{eq:timevolinhh}) implies that if we have $B_H$ time-independent then for any $v \in \hilbh$
\begin{equation} \nonumber
v(t;s,x,y) = e^{t  \antid B_H } v(0;s,x,y)
\end{equation}
where $e^{ \antid B_H } $ is the exponential of the operator of applying the left convolution of $B_H$ and then applying $\antid$.

\begin{Thm} \label{thm:schroheis}
If we have a system with energy $B_H$ (assumed to be Hermitian) then for any state $v \in \hilbh$ and any observable $B$

\begin{equation} \nonumber
\frac{d}{dt} \langle B * v, v \rangle = \langle \ub{B}{B_H}{} * v , v \rangle.
\end{equation}
\end{Thm}

\begin{proof}
The result follows from the direct calculation:
\begin{eqnarray} \nonumber
\frac{d}{dt} \langle B * v (t) , v (t) \rangle
&=& \langle B * \frac{d}{dt} v , v \rangle + \langle B * v, \frac{d}{dt} v \rangle \\ \nonumber
&=& \langle B * \antid B_H *v  , v \rangle + \langle B * v, \antid B_H *v \rangle  \\ \label{eq:usedadjofantid}
&=& \langle B* \antid B_H * v, v \rangle - \langle \antid B * v , B_H *v \rangle \\ \label{eq:usedhermitian}
&=& \langle B * \antid B_H * v  , v \rangle - \langle \antid  B_H * B * v, v \rangle  \\ \nonumber
&=& \langle \ub{B}{B_H}{} *v , v \rangle.
\end{eqnarray}
Equation (\ref{eq:usedadjofantid}) follows since $\antid$ is skew-adjoint in $\hilbh$. At (\ref{eq:usedhermitian}) we have used the fact that $B_H$ is Hermitian.
\end{proof}
This Theorem proves that the time evolution of states in $\hilbh$ coincides with the time evolution of observables as described in equation (\ref{eq:pdyneqn}). We now give a corollary to show that the time evolution of p-mechanical states in $\hilbh$, $h \in \Space{R}{} \setminus \{ 0 \} $ is the same as the time evolution of quantum states.
\begin{Corollary}
If we have a system with energy $B_H$ (assumed to be Hermitian) and an arbitrary state $v = \statem f = e^{2 \pi ihs} \hat{f} (x,y)$ (assuming $h \neq 0$) then for any observable $B (t;s,x,y)$
\begin{equation} \nonumber
\Diffl{t} \langle B * v (t) , v (t) \rangle_{\hilbh} = \Diffl{t} \langle \rho_h (B) f(t) , f(t) \rangle_{\fock}.
\end{equation}
Where $\Fracdiffl{f}{t} = \frac{1}{ih} \rho_{h} (B_H) f$ (this is just the usual Schr\"odinger equation).

\end{Corollary}
\begin{proof}
From Theorem \ref{thm:schroheis} we have
\begin{eqnarray} \nonumber
\frac{d}{dt} \langle  B * v, v \rangle &=&  \langle \ub{B}{B_H}{} * v , v \rangle \\ \nonumber
&=& \langle \antid (B * B_H - B_H * B ) * v, v \rangle \\ \nonumber
&=& \langle (B * B_H - B_H * B ) * \antid v, v \rangle \\ \nonumber
&=& \frac{2 \pi}{ih} \langle (B * B_H - B_H * B ) * v, v \rangle \\ \nonumber
&=& \frac{1}{i \hbar} ( \langle B * B_H *v,v \rangle - \langle B * v, B_H * v \rangle)
\end{eqnarray}
The last step follows since $B_H$ is Hermitian. Using equation (\ref{eq:statesrelation}), the above equation becomes,
\begin{eqnarray} \nonumber
\frac{d}{dt} \langle B * v, v \rangle &=& \frac{1}{i\hbar} (\langle \rho_h (B) \rho_h (B_H)f,f \rangle_{\fock} - \langle \rho_h (B) f , \rho_h (B_H )f \rangle_{\fock} )\\ \nonumber
&=& \frac{d}{dt} \langle \rho_h (B)f, f \rangle_{\fock},
\end{eqnarray}
which completes the proof.
\end{proof}

Hence the time development in $\hilbh$ for $h \neq 0$ gives the same time development as in $\fock$.

If $l(s,x,y) = \left( \frac{4}{h} \right)^n \int_{\Heisn} \overline{v((s',x',y'))} v((s',x',y')^{-1} (s,x,y)) \, dx' \, dy'$ then by Theorems \ref{Thm:timeevolofkernelisok} and \ref{thm:schroheis} we have that
\begin{equation} \label{eq:sametimeevolofkernelandvector}
\Diffl{t} \langle B*v,v \rangle_{\hilbh} = \Diffl{t} \int_{\Heisn} B \, l \, dg.
\end{equation}

\subsection{Eigenvalues and Eigenfunctions} \label{sect:eigenvalueeigenfunction}

In this section we introduce the concept of eigenvalues and eigenfunctions for  p-observables.

\begin{Thm}

For a p-observable $B \in \loneh$ and $f_1 \in \fock$, $\rho_h (B) f_1 = \lambda f_1$, if and only if for $v_1 (s,x,y) = \statem f_1 = e^{2 \pi ihs} \hat{f_1} (x,y) \in \hilbh$
\begin{equation} \nonumber
\langle B * v_1, v_2 \rangle = \lambda \langle v_1, v_2 \rangle
\end{equation}
holds for all $v_2 \in \hilbh$.
\end{Thm}

\begin{proof}

If $v_2 = e^{2\pi ihs} \hat{f_2} (x,y)$ where $f_2$ is an arbitary element of $\fock$.
\newline
$\rho_h (B) f_1 = \lambda f_1$ implies that
\begin{equation} \label{eiginf}
\langle \rho_h (B) f_1 ,f_2 \rangle =  \lambda \langle f_1,f_2 \rangle = \lambda \langle \rho_h (\zerodel) f_1 , f_2 \rangle
\end{equation}
and by (\ref{eq:statesrelation}) this gives us
\begin{equation} \label{eigink}
\langle B * v_1 , v_2 \rangle = \lambda \langle \zerodel * v_1, v_2 \rangle = \lambda \langle v_1 , v_2 \rangle.
\end{equation}
Which proves the argument in one direction. Clearly equations (\ref{eiginf}) and (\ref{eigink}) are equivalent so the converse follows since (\ref{eigink}) holding for any $v_2 \in \hilbh$ is equivalent to (\ref{eiginf}) holding for any $f_2 \in \fock$.
\end{proof}

\subsection{Coherent States} \label{sect:cstates}

In this section we introduce an overcomplete system of vectors in $\hilbh$ by a representation of $\Heisn$. The states which correspond to these vectors are an overcomplete system of coherent states for each $h \neq 0$. We then show that these vectors correspond to a system of kernels in $\lkerh$, whose limit is the $(q,p)$ pure state kernels.

Initally we need to introduce a vacuum vector in $\hilbh$. For this we take the vector in $\hilbh$ corresponding to the ground state of the Harmonic Oscillator with classical Hamiltonian $\frac{1}{2} (m \omega^2 q^2 + \frac{1}{m}p^2)$ where $\omega$ is the constant frequency and $m$ is the constant mass. The vector in $\fock$ corresponding to the ground state is \cite[Eq 2.18]{Kisil02}
\begin{equation} \nonumber
f_0 (q,p) = \exp \left( -\frac{2\pi}{h} (\omega m q^2 + (\omega m)^{-1} p^2 ) \right), \hspace{1cm} h>0 .
\end{equation}
The image of this under $\statem$ is
\begin{equation} \nonumber
e^{2 \pi ihs} \fort ( f_0 ) = e^{2 \pi ihs} \int_{\Space{R}{2n}}  e^{-\frac{2 \pi}{h} (m \omega q^2 + (m \omega )^{-1} p^2)} e^{-2 \pi i(qx+py)} \, dq \, dp.
\end{equation}
Using the basic formula
\begin{equation} \label{eq:thewaveletformula}
\int_{\Space{R}{}} \exp(-a x^2 + b x + c) dx = \left( \frac{\pi}{a} \right)^{\frac{1}{2}} \exp \left( \frac{b^2}{4a} + c \right), \textrm{   where  } a > 0
\end{equation}
we get
\begin{equation} \nonumber
\statem (f_0 ) =  e^{2 \pi ihs} \fort ( f_0 ) = \left( \frac{h}{2} \right)^n  \exp \left(  2 \pi ihs - \frac{\pi h}{2} \left( \frac{x^2}{\omega m} + y^2 \omega m \right) \right),
\end{equation}
which is the element of $\hilbh$ corresponding to the ground state.

\begin{Defn}
Define the vacuum vector in $\hilbh$ as
\begin{equation} \nonumber
\voo = \left( \frac{h}{2} \right)^n \exp \left( 2\pi i hs - \frac{\pi h}{2} \left( \frac{x^2}{\omega m} + y^2 \omega m \right) \right),
\end{equation}
where $\omega$ and $m$ are constants representing frequency and mass respectively.
\end{Defn}
Now we calculate the kernel, $\loo$, for the ground state by the relationship (\ref{eq:relationbetweenkernelandvector}) between kernels and vectors.

\begin{eqnarray} \nonumber
\lefteqn{\loo (s,x,y) } \\ \nonumber
&=& \left(\frac{4}{h} \right)^n \int_{\Space{R}{2n}} \voo ((-s,-x,-y)(s',x',y')) \overline{\voo(s',x',y')} \, dx' \, dy' \\ \nonumber
&=& h^n e^{-2\pi i hs} \int_{\Space{R}{2n}} \exp \left(\pi i h(x'y-xy') -\frac{\pi h}{2} \left(\frac{(x'-x)^2}{\omega m} + \omega m (y-y')^2  \right) \right.\\ \nonumber
&& \qquad  \qquad \qquad \qquad \qquad \left. - \frac{\pi h}{2} \left( \frac{(x')^2}{\omega m} + \omega m (y')^2  \right) \right) \, dx' \, dy' \\ \nonumber
&=& h^n \exp \left( -2\pi ihs -\frac{\pi h}{2} \left( \frac{x^2}{\omega m} +\omega m y^2 \right) \right) \\ \nonumber
&& \qquad \qquad \qquad \times \int_{\Space{R}{2n}} \exp \left(\pi h \left(-\frac{(x')^2}{\omega m}   + \left( iy+\frac{x}{\omega m} \right)x' \right. \right. \\ \nonumber
&& \qquad \qquad \qquad \qquad \qquad \qquad \qquad \left. \left. - \omega m (y')^2 + (\omega m y - ix )y' \right) \right) \, dx' \, dy'  \\ \label{eq:usedthewaveletformula}
&=& \exp \left( -2\pi ihs -\frac{\pi h}{2}(\frac{x^2}{\omega m} +\omega m y^2) \right) \\ \nonumber
&& \qquad \times \exp \left(\frac{\pi h}{4} \left( \omega m \left(iy+\frac{x}{\omega m}\right)^2 + \frac{1}{\omega m} ( \omega m y -ix)^2 \right) \right)
\end{eqnarray}
at (\ref{eq:usedthewaveletformula}) we have used formula (\ref{eq:thewaveletformula}). By a simple calculation it can be shown that
\begin{equation} \nonumber
\omega m \left( iy+\frac{x}{\omega m}\right)^2 + \frac{1}{\omega m} (y \omega m -ix)^2  = 0
\end{equation}
hence
\begin{equation} \nonumber
\loo = \exp \left( -2\pi ihs -\frac{\pi h}{2}\left( \frac{x^2}{\omega m} +\omega m y^2 \right) \right)
\end{equation}
From \cite{Kisil02.1} we introduce the observables $X$ and $Y$, which are convolutions with the following distributions
\begin{displaymath}
\begin{array}{ccc}
X= \frac{1}{2 \pi i} \zerodelxone & \textrm{ and } & Y = \frac{1}{2 \pi i} \zerodelyone.
\end{array}
\end{displaymath}
Under left and right convolution $X$ and $Y$ generate left and right invariant vector fields respectively. That is, if $B$ is a function or distribution on $\Heisn$ then
\begin{displaymath}
\begin{array}{ccc} \nonumber
X*B = \frac{1}{2 \pi i}(\Partial{x} - \frac{y}{2} \Partial{s})B & \hspace{1cm} & B*X = \frac{1}{2 \pi i}(\Partial{x} + \frac{y}{2} \Partial{s})B \\ \nonumber
Y*B = \frac{1}{2 \pi i}(\Partial{y} + \frac{x}{2} \Partial{s})B & \hspace{1cm} & B*Y = \frac{1}{2 \pi i}(\Partial{y} - \frac{x}{2} \Partial{s})B
\end{array}
\end{displaymath}

Consider the representation of $\Heisn$ on $\hilbh$ by
\begin{equation} \nonumber
\zeta_{(r,a,b)} v(s,x,y) = e^{-2 \pi i r s} e^{- 2 \pi i \antid (-b X  + a Y)} v(s,x,y),
\end{equation}
where $e^X$ is exponential of the operator of convolution by $X$.
The elements $(r,0,0)$ act trivally in the representation, $\zeta$, thus the essential part of the operator $\zeta_{(r,a,b)}$ is determined by $(a,b)$. Physically the $e^{-2\pi irs}$ part of the equation will just be a phase factor which can be ignored. If we apply this representation with $r=0$ to $\voo$ we get a system of vectors $\vab$,
\begin{equation} \nonumber
v_{(h,a,b)} (s,x,y) = \zeta_{(0,a,b)} \left( \left( \frac{h}{2} \right)^n   \exp \left( 2 \pi ihs \frac{- \pi h}{2} \left( \frac{x^2}{\omega m} + y^2 \omega m \right) \right) \right).
\end{equation}

By (\ref{eq:sametimeevolofkernelandvector}) the vectors $\vab$ are equivalent to the kernels $\lab$
\begin{equation} \nonumber
\lab = e^{2\pi i (-b \ub{X}{\cdot}{} + a \ub{Y}{\cdot}{} )} \loo.
\end{equation}
Since for any function or distribution, $B$, on $\Heisn$
\begin{equation} \nonumber
\ub{ -b X + a Y  }{B} = -(a x + b y )B
\end{equation}
we have
\begin{equation} \nonumber
\lab =  \exp \left( -2\pi i (a \cdot x + b \cdot y) -2\pi ihs -\frac{\pi h}{2} \left( \frac{x^2}{\omega m} +\omega m y^2 \right) \right).
\end{equation}
\begin{Defn}
For $h \in \Space{R}{} \setminus \{ 0 \}$ and $(a,b) \in \Space{R}{2n}$ define the system of coherent states $k_{(h,a,b)}$ by
\begin{equation} \nonumber
k_{(h,a,b)} (B) = \langle B* \vab,\vab \rangle = \int_{\Heisn} B(g) \lab (g) dg
\end{equation}
\end{Defn}
It is clear that the limit as $h \rightarrow 0$ of the kernels $\lab$ will just be the kernels  $\labo$. This proves that the system of coherent states we have constructed have the $(q,p)$ pure states, $k_{(0,a,b)}$, from equation (\ref{eq:pclasspurestates}) , as their limit as $h \rightarrow 0$, which is the content of the next Theorem.
\begin{Thm}
If we have any p-observable $B$ which is of the form $\delta(s) \hat{F}(x,y)
\newline
$(that is, $B$ is the p-mechanisation of $F$ as described in \emph{\cite[Sect. 3.3]{Kisil02.1}}) then
\begin{equation} \nonumber
\lim_{h \rightarrow 0} k{(h,a,b)} (B) = k{(0,a,b)} (B) = F(a,b)
\end{equation}
\end{Thm}
We have used p-mechanics to rigorously prove, in a simpler way to previous attempts \cite{Hepp74}, the classical limit of coherent states.

\subsection{The Interaction Picture} \label{sect:interactpic}

In the Schr\"odinger picture, time evolution is governed by the states and their equations $\frac{d v}{dt} = \antid B_H * v$ $\Fracdiffl{l}{t}= \ub{B_H}{l}{}$. In the Heisenberg picture, time evolution is governed by the observables and the equation $\frac{d B}{dt} = \ub{B}{B_H}{}$. In the interaction picture we divide the time dependence between the states and the observables. This is suitable for systems with a Hamiltonian of the form $B_H = B_{H_0} + B_{H_1}$ where $B_{H_0}$ is time independent. The interaction picture has many uses in perturbation theory \cite{Kurunoglu62}.

Let a p-mechanical system have the Hamiltonian $B_H = B_{H_0} + B_{H_1}$ where $B_{H_0}$ is time independent. We first describe the interaction picture for elements of $\hilbh$. Define  $\exp (t \antid  B_{H_0})$ as the operator on $\hilbh$ which is the exponential of the operator of convolution by $t \antid B_{H_0}$. Now if $B$ is an observable let
\begin{equation} \label{eq:obsevolveininter}
\tilde{B} = \exp(t \antid B_{H_0}) B \exp(-t \antid B_{H_0})
\end{equation}
If $v \in \hilbh$, define $\tilde{v} = (\exp (- t \antid B_{H_0} )) v$, then we get
\begin{eqnarray} \label{eq:gentimeevolofinter}
\frac{d}{dt} \tilde{v} &=& \frac{d}{dt} ( \exp(-t \antid B_{H_0}) v ) \\ \nonumber
&=& - \antid B_{H_0} * \tilde{v} + \exp(-t \antid B_{H_0}) ( \antid (B_{H_0} + B_{H_1}) *v) \\ \nonumber
&=& - \antid B_{H_0} * \tilde{v} + \antid B_{H_0} * \exp(-t \antid B_{H_0}) v + \exp(-t \antid B_{H_0}) \antid B_{H_1} v \\ \nonumber
&=& (\exp(-t \antid B_{H_0}) \antid B_{H_1} \exp(t \antid B_{H_0}))(\tilde{v})
\end{eqnarray}
Now we describe the interaction picture for a state defined by a kernel $l$. Define
\begin{equation} \nonumber
\tilde{l} = e^{\ub{B_{H_0}}{\cdot}{}t} l =  \exp(t \antid B_{H_0}) l \exp(-t \antid B_{H_0})
\end{equation}
then
\begin{eqnarray} \nonumber
\Fracdiffl{\tilde{l}}{t} &=& \antid B_{H_0} * \tilde{l} + \exp(t \antid B_{H_0}) \ub{B_{H_0} + B_{H_1}}{l}{} \exp(-t \antid B_{H_0}) - \tilde{l} * \antid B_{H_0} \\ \nonumber
&=& \exp(t \antid B_{H_0}) \ub{B_{H_1}}{l}{} \exp(-t \antid B_{H_0}) \\ \nonumber
&=& \exp(t \antid B_{H_0}) (\antid (B_{H_1} * \exp(-t \antid B_{H_0}) \tilde{l} \exp(t \antid B_{H_0}) \\ \nonumber
&& \qquad - \exp(-t \antid B_{H_0}) \tilde{l} \exp(t \antid B_{H_0}) * B_{H_1} )) \exp(-t \antid B_{H_0}) \\ \nonumber
&=& \ub{\exp(t \antid B_{H_0}) B_{H_1} \exp(-t \antid B_{H_0})} {\tilde{l}}{}
\end{eqnarray}
This shows us how interaction states evolve with time, while the observables evolve by (\ref{eq:obsevolveininter}). Note that if we take $B_{H_0}=B_H$ we have the Heisenberg picture, while if we take $B_{H_1} = B_H$ we have the Schr\"odinger picture. The interaction picture is very useful in studying the forced harmonic oscillator as will be shown in subsection \ref{sect:interpictofforcedosc}.

\section{The Forced Harmonic Oscillator} \label{sect:forcedosc}

The classical forced oscillator has been studied in great depth for a long time --- for a description of this see \cite{Jose98} and \cite{Goldstein80}. The quantum case has also been heavily researched --- see for example \cite[Sect 14.6]{Merzbacher70}, \cite{Martinez83}. Of interest in the quantum case has been the use of coherent states, this is described in \cite{Perelomov86}. Here we extend these approaches to give a unified quantum and classical solution of the problem based on the p-mechanical approach.

\subsection{The Unforced Harmonic Oscillator}

Initially we give a brief overview of the unforced harmonic oscillator; we give a slightly different account to the one given in \cite{Kisil02.1}.
\begin{Defn}
We define the p-mechanical creation and annihilation operators respectively as convolution by the following distributions
\begin{eqnarray} \label{eq:defaplus}
a^+ &=& \frac{1}{2\pi i} (m \omega \zerodelxone - i  \zerodelyone), \\ \label{eq:defaminus}
a^- &=& \frac{1}{2 \pi i} (m \omega \zerodelxone + i  \zerodelyone).
\end{eqnarray}
\end{Defn}
The p-mechanical harmonic oscillator Hamiltonian has the equivalent form
\begin{equation} \nonumber
B_H = \frac{1}{2m} (a^+ * a^- + i \omega m^2 \zerodelsone ).
\end{equation}
We denote the p-mechanical normalised eigenfunctions of the harmonic oscillator by $v_n \in \hilbh$ (note here that $v_{(h,0,0)}=v_0$); they have the form
\begin{eqnarray} \nonumber
v_n &=& \left( \frac{1}{n!} \right)^{1/2} (\antid a^+)^n * v_{(h,0,0)} \\ \nonumber
&=& \left( \frac{1}{n!} \right)^{1/2} \left( \frac{h}{2} \right)^n e^{2\pi i hs} (x+i\omega m y)^n  \exp \left( \frac{- \pi h}{2} \left( \frac{x^2}{\omega m} + y^2 \omega m \right) \right).
\end{eqnarray}
It can be shown by a trivial calcuation that these creation and annihilation operators raise and lower the eigenfunctions of the harmonic oscillator respectively. It is important to note that these states are orthogonal under the $\hilbh$ inner product defined in equation (\ref{hhip}).

\subsection{The p-Mechanical Forced Oscillator: The Solution and Relation to Classical Mechanics} \label{sect:forcedoscclass}

The classical Hamiltonian for a Harmonic oscillator of frequency $\omega$ and mass $m$ being forced by a real function of a real variable $z(t)$ is
\begin{equation} \nonumber
H(t,q,p) = \frac{1}{2} \left( m \omega^2 q^2 + \frac{1}{m} p^2 \right) - z(t) q.
\end{equation}
Then for any observable $f \in C^{\infty} (\Space{R}{2n})$ the dynamic equation is
\begin{eqnarray} \nonumber
\Fracdiffl{f}{t} &=& \{ f,H \} \\ \label{eq:classdyneqnfo}
&=& \frac{p}{m} \Fracpartial{f}{q} - \omega^2 m q \Fracpartial{f}{p} + z(t) \Fracpartial{f}{p}.
\end{eqnarray}
Through the procedure of p-mechanisation as described in \cite[Sect 3.3]{Kisil02.1} we get the p-mechanical forced oscillator Hamiltonian to be
\begin{eqnarray} \nonumber
B_H (t;s,x,y) &=& -\frac{1}{8 \pi^2} \left( m \omega^2 \zerodelxtwo + \frac{1}{m} \zerodelytwo \right) \\ \nonumber
&& \qquad- \frac{z(t)}{2 \pi i} \zerodelxone.
\end{eqnarray}
From equation (\ref{eq:pdyneqn}) the dynamic equation for an arbitary observable $B$ is
\begin{equation} \label{eq:pdynamiceqnforforcedosc}
\Fracdiffl{B}{t} = \frac{x}{m} \Fracpartial{B}{y} - \omega^2 m y \Fracpartial{B}{x} - z(t) y B.
\end{equation}
By substitutiting the following expression into equation (\ref{eq:pdynamiceqnforforcedosc}) we see that it is a solution of the p-dynamic equation
\begin{eqnarray} \label{eq:solnofpmfo}
\lefteqn{B (t;s,x,y)} \\ \nonumber
&& = \exp \left( 2 \pi i \left( \frac{1}{m \omega} \int_{0}^{t} z(\tau) \sin (\omega \tau ) \, d\tau X(t) - \int_{0}^{t} z(\tau ) \cos (\omega \tau ) \, d \tau Y(t) \right) \right) \\ \nonumber
&& \qquad \qquad \times B (0;s,X(t),Y(t)),
\end{eqnarray}
where
\begin{eqnarray} \nonumber
X(t) &=& x \cos(\omega t) - m \omega y \sin (\omega t), \\ \nonumber
Y(t) &=& \frac{x}{m\omega} \sin (\omega t) + y \cos (\omega t).
\end{eqnarray}
Let $F (q,p) = \rho_{(q,p)}(B (s,x,y))$ (i.e. $F$ is the classical observable corresponding to $B$ under the relationship described in \cite[Sect. 3.3]{Kisil02.1}).
\begin{eqnarray} \nonumber
\lefteqn{ F(t;q,p) } \\ \nonumber
&& = \int_{\Space{R}{2n+1}} B (t;s,x,y) e^{2\pi i (q.x+p.y)} \, ds \, dx \, dy  \\ \nonumber
&& = \int_{\Space{R}{2n+1}} \exp \left( 2 \pi i \left( \frac{1}{m \omega} \int_{0}^{t} z(\tau) \sin (\omega \tau ) \, d\tau X(t) - \int_{0}^{t} z(\tau ) \cos (\omega \tau ) \, d \tau Y(t) \right) \right) \\ \nonumber
&& \qquad \qquad \times  \exp (2\pi i (q.x +p.y) ) \, B(0;s,X(t),Y(t)) \, ds \, dx \, dy.
\end{eqnarray}
Making the change of variable $u=X(t)$ and $v=Y(t)$ the above equation becomes
\begin{eqnarray} \nonumber
\lefteqn{\int_{\Space{R}{2n+1}} \exp \left( 2 \pi i \left( \frac{1}{m \omega} \int_{0}^{t} z(\tau) \sin (\omega \tau ) \, d\tau u - \int_{0}^{t} z(\tau ) \cos (\omega \tau ) \, d \tau v \right) \right) } \\ \nonumber
&& \times \exp \left(2\pi i \left( q.(u \cos(\omega t) +vm\omega \sin(\omega t)) +p.(-\frac{u}{m\omega} \sin(\omega t) +v \cos(\omega t)) \right) \right) \\ \nonumber
&& \qquad \qquad \times B (0;s,u,v) \, ds \, du \, dv \\ \nonumber
&& = \int_{\Space{R}{2n+1}} \exp \left( 2\pi i u. \left( q cos(\omega t) -\frac{p}{m\omega} \sin (\omega t) + \frac{1}{m \omega} \int_{0}^{t} z(\tau) \sin (\omega \tau ) \, d\tau \right) \right) \\ \nonumber
&& \qquad \times \exp \left( 2 \pi i v. \left( q m\omega \sin (\omega t) +p \cos (\omega t) - \int_{0}^{t} z(\tau ) \cos (\omega \tau ) \, d \tau \right) \right) \\ \nonumber
&& \qquad \qquad \times B (0;s,u,v) \, ds \, du \, dv \\ \nonumber
&& = F \left( 0; q \cos(\omega t) -\frac{p}{m\omega} \sin (\omega t) + \frac{1}{m \omega} \int_{0}^{t} z(\tau) \sin (\omega \tau ) \, d\tau , \right.
\\ \label{eq:classicalflowfo}
&& \left. \qquad \qquad q m\omega \sin (\omega t) +p \cos (\omega t) - \int_{0}^{t} z(\tau ) \cos (\omega \tau ) \, d \tau \right).
\end{eqnarray}
This flow satisfies the classical dynamic equation (\ref{eq:classdyneqnfo}) for the forced oscillator --- this is shown in \cite{Jose98}.

\subsection{A Periodic Force and Resonance} \label{sect:periodandres}

In classical mechanics the forced oscillator is of particular interest if we take the external force to be $z(t)=Z_0 \cos(\Omega t)$ \cite{Jose98}, that is the oscillator is being driven by a harmonic force of constant frequency $\Omega$ and constant amplitude $Z_0$. By a simple calculation we have these results for $\Omega \neq \omega$
\begin{eqnarray} \nonumber
\lefteqn{\int_0^t \cos (\Omega \tau) \sin(\omega \tau) \, d \tau} \\ \label{eq:sinint}
&& \qquad = \frac{2}{(\Omega^2 -\omega^2 ) } [ \Omega \cos (\Omega t) \cos (\omega t) + \omega \sin (\Omega t) \sin (\omega t) - \Omega ]
\end{eqnarray}
\begin{eqnarray} \nonumber
\lefteqn{ \int_0^t \cos (\Omega \tau) \cos(\omega \tau) \, d \tau} \\ \label{eq:cosint}
&& = \frac{2}{(\Omega^2 -\omega^2 ) } [ -\Omega \sin (\Omega t) \cos (\omega t) + \omega \cos (\Omega t) \sin (\omega t) ]
\end{eqnarray}

When these are substituted into (\ref{eq:solnofpmfo}) we see that in p-mechanics using a periodic force the p-mechanical solution is the flow of the unforced oscillator multiplied by an exponential term which is also periodic. However this exponential term becomes infinitely large as $\Omega$ comes close to $\omega$. If we substitute $(\ref{eq:sinint})$ and $(\ref{eq:cosint})$ into $(\ref{eq:classicalflowfo})$ we obtain a classical flow which is periodic but with a singularity as $\Omega$ tends towards $\omega$. These two effects show a correspondence between classical and p-mechanics. The integrals have a different form when $\Omega = \omega$

\begin{eqnarray} \label{eq:sinint2}
\int_0^t \cos (\omega \tau) \sin(\omega \tau) \, d \tau &=& \frac{1-\cos(2\omega t)}{4 \omega} \\ \label{eq:cosint2}
\int_0^t \cos (\omega \tau) \cos(\omega \tau) \, d \tau &=& \frac{t}{2} + \frac{1}{4 \omega} \sin(2\omega t)
\end{eqnarray}
Now when these new values are substituted into the p-mechanical solution (\ref{eq:solnofpmfo}) the exponential term will expand without bound as $t$ becomes large. When (\ref{eq:sinint2}) and (\ref{eq:cosint2}) are substituted into $(\ref{eq:classicalflowfo})$ the classical flow will also expand without bound --- this is the effect of resonance.

\subsection{The Interaction Picture of the Forced Oscillator} \label{sect:interpictofforcedosc}

We now use the interaction picture to get a better description of the p-mechanical forced oscillator and also to demonstrate some of the quantum effects. The method we use is similar to the known method for the quantum system \cite{Merzbacher70}. The p-mechanical forced oscillator Hamiltonian has the equivalent form
\begin{equation} \nonumber
B_H = \frac{1}{2m} \left( a^+ * a^- + i \omega m^2 \zerodelsone \right) - z(t) (a^- + a^+ )
\end{equation}
($a^+$ and $a^-$ are the distributions defined in equations (\ref{eq:defaplus}) and (\ref{eq:defaminus})).
We now proceed to solve the Forced Oscillator in p-mechanics using the interaction picture with $B_{H_0} = \frac{1}{2m} (a^+ * a^- + i \omega m^2 \zerodelsone )$ and $B_{H_1} =  - z(t) (a^- + a^+ )$. From (\ref{eq:gentimeevolofinter}) the interaction states evolve under the equation
\begin{eqnarray} \label{eq:fointerhamil}
\Fracdiffl{\tilde{v}}{t} &=& \exp \left( \frac{t}{2m} \antid (a^+ * a^- + i \omega m^2 \zerodelsone ) \right) \\ \nonumber
&& \times ( - \antid z(t) (a^- + a^+ )) \exp \left( -\frac{t}{2m} \antid  (a^+ * a^- + i \omega m^2 \zerodelsone ) \right) \tilde{v},
\end{eqnarray}
where $\tilde{v} = e^{t \antid B_0} v$ and the exponentials are exponentials of the operators of convolution by the appropriate distributions.
\begin{Lemma} \label{lemma:anihilrel}
We have the relations
\begin{eqnarray} \label{eq:firstanihilrel}
\ub{a^+}{a^-} &=& i \omega m \zerodel \\ \label{eq:secondanihilrel}
\ub{a^+}{a^+ * a^-} &=& i \omega m a^+ \\ \label{eq:thirdanihilrel}
\ub{a^-}{a^+ * a^-} &=& -i \omega m a^- .
\end{eqnarray}
\end{Lemma}
\begin{proof}
Equation (\ref{eq:firstanihilrel}) follows from simple properties of commutation for convolutions of Dirac delta functions. Equations (\ref{eq:secondanihilrel}) and (\ref{eq:thirdanihilrel}) follow from (\ref{eq:firstanihilrel}) and  the fact that $\ub{}{}$ are a derivation.
\end{proof}

\begin{Lemma} \label{lemma:expcommidentity}
If $B_1,B_2$ are functions or distributions on $\Heisn$ such that
\newline
$\ub{B_1}{B_2}{} = \gamma B_2$ where $\gamma$ is a constant then we have
\begin{equation} \label{eq:expcommidentity}
e^{\antid \lambda B_1} \antid B_2 e^{-\antid \lambda B_1} = e^{\lambda \gamma} \antid B_2.
\end{equation}
Here $e^{\antid \lambda B_1}$ is the exponential of the operator of convolution by $\antid \lambda B_1$.

\end{Lemma}
\begin{proof}
It is clear that
\begin{eqnarray} \nonumber
[\antid B_1, \antid  B_2] &=& \antid \ub{B_1}{B_2}{} \\ \nonumber
&=& \gamma \antid B_2.
\end{eqnarray}
We have the operator identity \cite[Eq 3.59]{Merzbacher70}: if $[C_1,C_2] = \gamma C_2$ then
\newline
$e^{\lambda C_1} C_2 e^{-\lambda C_1} = e^{\lambda \gamma} C_2$. Equation (\ref{eq:expcommidentity}) is  this with $C_1$ and $C_2$ the operators  $\antid B_1$  and $\antid B_2$ respectively.
\end{proof}
The combination of Lemmas \ref{lemma:anihilrel} and \ref{lemma:expcommidentity} simplifies equation (\ref{eq:fointerhamil}) to
\begin{eqnarray} \nonumber
\Fracdiffl{\tilde{v}}{t} &=& -\antid  z(t) (a^- e^{i\omega t} + a^+ e^{-i \omega t}) \tilde{v} \\ \nonumber
&=& -\antid  (f(t) a^- + \overline{f(t)} a^+) \tilde{v}
\end{eqnarray}
where $f(t) = z(t) e^{i\omega t}= z(t) \cos (\omega t) + i z(t) \sin (\omega t)$.
A solution of this is
\begin{eqnarray} \nonumber
\lefteqn{\tilde{v}(t_2;s,x,y) = \exp \left( \antid \left( \int_{t_1}^{t_2} a^- f(\tau)  + a^+ \overline{f(\tau)} \, d \tau  \right.  \right. } \\ \label{eq:timedeloffo}
&& \left. \left. + \omega m \int_{t_1}^{t_2} \int_{t_1}^{t_2} z(\tau) z(\tau') cos (\omega (\tau - \tau'))) \, d\tau \, d\tau' \right) \right) \tilde{v} (t_1;s,x,y)
\end{eqnarray}
If we set
\begin{eqnarray} \nonumber
\xi (t_2,t_1) &=&  \omega m \int_{t_1}^{t_2} \int_{t_1}^{t_2} z(\tau) z(\tau ') cos(\omega (\tau - \tau ' ))\, d \tau \, d \tau ' \\ \nonumber
\eta(t_2,t_1) &=& \int_{t_1}^{t_2} f(\tau)  \, d \tau.
\end{eqnarray}
equation (\ref{eq:timedeloffo}) becomes
\begin{eqnarray} \nonumber
\tilde{v}(t_2;s,x,y) &=& e^{\antid \xi (t_2,t_1)} e^{\antid (\bar{\eta} (t_2,t_1) a^{+} + \eta (t_2 ,t_1 ) a^{-} )} \tilde{v} (t_1;s,x,y) \\ \nonumber
&=& e^{\antid \xi (t_2,t_1)} e^{\antid (\omega m \eta_R (t_2,t_1) X - \eta_I (t_2,t_1) Y) } \tilde{v} (t_1;s.x.y),
\end{eqnarray}
where $\eta_R (t_2,t_1)$ and $\eta_I (t_2,t_1)$ are respectively the real and imaginary parts of $\eta ( t_2 , t_1 )$.
Hence if at time $t_1$ we start with a coherent state
$v_{(a,b)}$ this will evolve (in the interaction picture) to the state
\begin{equation} \label{eq:phasefactor}
e^{\xi (t_2,t_1) \antid } v_{(a + \omega m \eta_R (t_2,t_1),b - \eta_I (t_2,t_1))}.
\end{equation}
The $e^{\xi (t_2,t_1) \antid }$ part of formula (\ref{eq:phasefactor}) is just a phase factor and will be dealt with in the next subsection.
Observables in the interaction picture will evolve by equation (\ref{eq:obsevolveininter}) which is just the time evolution of observables for the unforced Harmonic oscillator. From \cite{Kisil02} this is
\begin{eqnarray} \nonumber
B(t_2;s,x,y) &=& B(t_1;s,  x cos(\omega (t_2-t_1)) - m \omega y sin (\omega (t_2-t_1)), \\ \nonumber
&& \qquad \frac{x}{m\omega} sin (\omega (t_2 - t_1)) + y cos (\omega (t_2 - t_1))).
\end{eqnarray}
{\bf Remark:} The states remaining coherent means if we let $h \rightarrow 0$ we can consider the classical time evolution by evaluating the observables at different points (that is the co-ordinates given by the coherent state). The observables themselves are moving, but just as they would under the unforced oscillator.
\subsection{The Quantum Case} \label{sect:forcedoscquant}

We define the time evolution operator (propagator), $T(t_2,t_1)$, of a system as
\begin{equation} \nonumber
v(t_2)= T(t_2,t_1)v(t_1)
\end{equation}
where $v \in \hilbh$ is a state evolving in the system. In the interaction picture of the Forced Harmonic Oscillator the time evolution operator is
\begin{equation} \nonumber
T(t_2,t_1) =  e^{\antid \xi (t_2,t_1)} e^{\antid (\bar{\eta} (t_2,t_1) a^{+} + \eta (t_2 ,t_1 ) a^{-}) }.
\end{equation}
The $S$-matrix (scattering matrix) in the interaction picture is something of particular interest \cite{Merzbacher70,Perelomov86}. The p-mechanical $S$-matrix in the interaction picture is
\begin{equation} \nonumber
S=T(\infty,-\infty)= e^{\xi_s \antid} e^{\antid (\eta_s a^- + \bar{\eta}_s a^+ )}
\end{equation}
where $\xi_s = \omega m \int_{\Space{R}{}}  \int_{\Space{R}{}} z(\tau) z(\tau ') \cos ( \omega (\tau - \tau')) \, d\tau \, d \tau '$ and $\eta_s = \int_{\Space{R}{}} z(\tau) e^{i \omega \tau} \, d \tau$. The function $\xi_s$ is well defined since in all cases the force can only act for a finite period of time. Since now we are dealing with only the quantum case we can assume that $h \neq 0$ and hence
\begin{equation} \nonumber
S=e^{\frac{2\pi \xi_s}{ih}} e^{\antid (\eta_s a^- + \bar{\eta}_s a^+ )}.
\end{equation}
The $e^{\frac{2\pi \xi_s}{ih}}$ is just a phase factor. We now introduce a well known operator identity which is a consequence of the Campbell-Baker-Hausdorff formula \cite{Merzbacher70}.
\begin{Lemma}
If $A_1$ and $A_2$ are two operators which commute with their commutator $[A_1,A_2]$ then
\begin{equation} \label{eq:expcommid1}
e^{A_1} e^{A_2} = e^{A_1 + A_2 + \frac{1}{2}[A_1,A_2]},
\end{equation}
and
\begin{equation} \label{eq:expcommid2}
e^{A_1} e^{A_2} e^{- \frac{1}{2}[A_1,A_2]} = e^{A_1 + A_2} .
\end{equation}
\end{Lemma}
For a proof of this Lemma see \cite[Chap 3]{Merzbacher70}. Using this Lemma we get
\begin{eqnarray} \nonumber
S &=&  e^{\frac{2\pi \xi_s}{ih}} e^{\antid \eta_s a^+} e^{\antid \bar{\eta}_s a^- } e^{ -\frac{1}{2}[  \antid \eta_s a^+ , \antid \bar{\eta}_s a^- ] } \\ \nonumber
&=& e^{\frac{2\pi \xi_s}{ih}} e^{\antid \eta_s a^+} e^{\antid \bar{\eta}_s a^- } e^{ - \frac{1}{2}| \eta_s |^2 i \omega m \antid }.
\end{eqnarray}
From this we can see if the oscillator begins in the oscillator state $v_0$ the probability amplitude of it being in the nth oscillator state $v_n$ is
\begin{eqnarray} \nonumber
| \langle S * v_{(0,0)} , v_n \rangle |^2
&=& | \langle  e^{\antid \eta_s a^+} e^{\antid \bar{\eta}_s a^-} e^{-\frac{| \eta_s |^2}{2} i \omega m \antid } v_{(0,0)},v_n \rangle |^2 \\ \nonumber
&=& | \langle e^{\antid \eta_s a^+} e^{\antid \bar{\eta}_s a^-} e^{-\frac{| \eta_s |^2}{2} i \omega m \frac{2 \pi}{i h} } v_{(0,0)},v_n \rangle |^2 \\ \nonumber
&=& | \langle e^{\frac{-\pi \omega m | \eta_s |^2}{h}  } e^{\antid \eta_s a^+} e^{\antid \bar{\eta}_s a^-} v_{(0,0)},v_n \rangle |^2 \\ \nonumber
&=& | \langle   e^{\frac{-\pi | \eta_s |^2  \omega m}{h}  } e^{\antid \eta_s a^+}  v_{(0,0)},v_n \rangle |^2 \\ \nonumber
&=& | \langle e^{\frac{ -\pi | \eta_s |^2  \omega m}{h}  }\sum_{j=1}^{\infty} \frac{(\antid \eta_s a^+)^j}{j!} v_{(0,0)}, \frac{(\antid a^+)^n}{(n!)^{1/2}} v_{(0,0)} \rangle |^2.
\end{eqnarray}
Using the orthonormality of the eigenstates this becomes
\begin{equation} \nonumber
| e^{\frac{-\pi | \eta_s |^2  \omega m}{i h}} \frac{(\frac{2 \pi}{i h} \eta_s )^n}{(n!)^{1/2}} |^2 = e^{-\frac{2\pi | \eta_s |^2  \omega m}{h}} \frac{ (\eta_s )^{2n}}{n!} = e^{-\frac{| \eta_s |^2  \omega m }{\hbar}} \frac{(\eta_s )^{2n}}{n!}.
\end{equation}
This is the same probability as can be found using normal quantum methods (there is a difference by $h$ compared to some of the literature but this is due to a different definition of $z(t)$ - see \cite[Eq. 14.107]{Merzbacher70} ).

\section*{Acknowledgements}
I would like to thank my supervisor Dr V. V. Kisil for all the help and encouragement he has given me while writing this paper. I would also like to thank Professor A.A. Kirillov for a useful comment.

\bibliography{everything}

\bibliographystyle{plain}

\end{document}